\documentclass[intlimits,twoside,a4paper]{article}

\usepackage[T2A]{fontenc} 
\usepackage[cp1251]{inputenc}

\usepackage[eqsecnum]{cmpj3}

\issue{2025}{28}{3}{33603}
\doinumber{10.5488/CMP.28.33603}

\title[Phase stability and structural properties]%
{Phase stability and structural properties of the K$_{x}$Ca$_{1-x}$N novel ferromagnetic alloy from first-principles}
\author[K. Larbaoui, A. Lakdja, G. Bassou]
{K. Larbaoui\orcid{0009-0002-0795-5275},
A. Lakdja\orcid{0000-0001-5403-5143}\thanks{Corresponding author: \email{abdelaziz.lakdja@univ-sba.dz}.},
G. Bassou\orcid{0000-0003-2985-2779}}
\address{
 Laboratoire de Microscopie, Microanalyse de la Mati\`{e}re et Spectroscopie Moleculaire,
Universit\'{e} Djillali Liab\`{e}s, 22000 Sidi-Bel-Abb\`{e}s, Algeria
}

\date{Received April 9, 2025, in final form June 23, 2025}

\begin{document}

\maketitle

\begin{abstract}
We study the structural properties and phase stability of the K$_{x}$Ca$_{1-x}$N alloy using the regular-solution model based on the total energy of the mixing. The pseudopotential approach was used along with PBE functional of Perdew, Burke, and Ernzerhof (PBE). We investigated the bond-lengths distribution as a function of composition~$x$. We also predicted the phase separation of the two partially miscible components and calculated the enthalpy $\Delta H$ using the interaction parameter $\Omega$. We observe an asymmetry about $x=0.46$ in the phase diagram due to the $x$-dependant interaction parameter $\Omega=12.69-1.32x$~kcal/mole. The equilibrium solubility limit, known as the miscibility gap is found to be around 3033~K.
%
%
\printkeywords
%
\end{abstract}

\section{Introduction}
Stimulated initially by developments in the field of high density magnetic recording, spintronics has become a growing discipline. In its development, several remarkable discoveries have opened up new fields of investigation and applications. Among these, the effects of tunnel magnetoresistance at room temperature and spin transfer were particularly significant steps. Dilute magnetic semiconductors (DMSs) was the best choice in this field \cite{Liu04,Gray12,Sanvito01}. The magnetic properties arise from their electronic states at the Fermi level in one spin channel which induce a high spin-polarized current \cite{Katsnelson08,Groot83}. With large spin-magnetic moments, the DMSs are expected to present large external stray magnetic fields and thus exhibit considerable energy losses. A serious alternative to the DMSs are systems where the magnetic order is carried by the $p$-electrons without any contribution of $d$-shell, namely $d^{0}$-half-metallic ferromagnets. The latter are based on alkali and alkaline earth metal elements, where the magnetic order is induced by the polarization of the anion $p$-empty states. Many studies on V- and VI-elements predicted half-metallicity with ferromagnetic order in various structures \cite{int01,int02,int03,int04,int05,int06,int07,int08,int09,int10,int11}. Moreover, recent studies confirm the phonons dynamical and mechanical stability of RbN, CsN, RbC, and SrC magnetic compounds \cite{phon1,phon2}.

On the order hand, novel alloys such as Rb$_{x}$Sr$_{1-x}$C and K$_{x}$Ca$_{1-x}$N exhibit an interesting behavior of magnetic order versus $x$ composition in the zinc-blende structure \cite{alloy1,alloy2}. In particular, the magnetic order changes from ferromagnetic to antiferromagnetic by varying the composition $x$ \cite{alloy1}. In the present paper we focused on the phase diagram of the K$_{x}$Ca$_{1-x}$N in its ferromagnetic state. We investigated the stable and unstable regions using the regular solution model from total energy calculation.

\section{Methodology}
\begin{figure}[h]
\centerline{\includegraphics[width=0.5\textwidth]{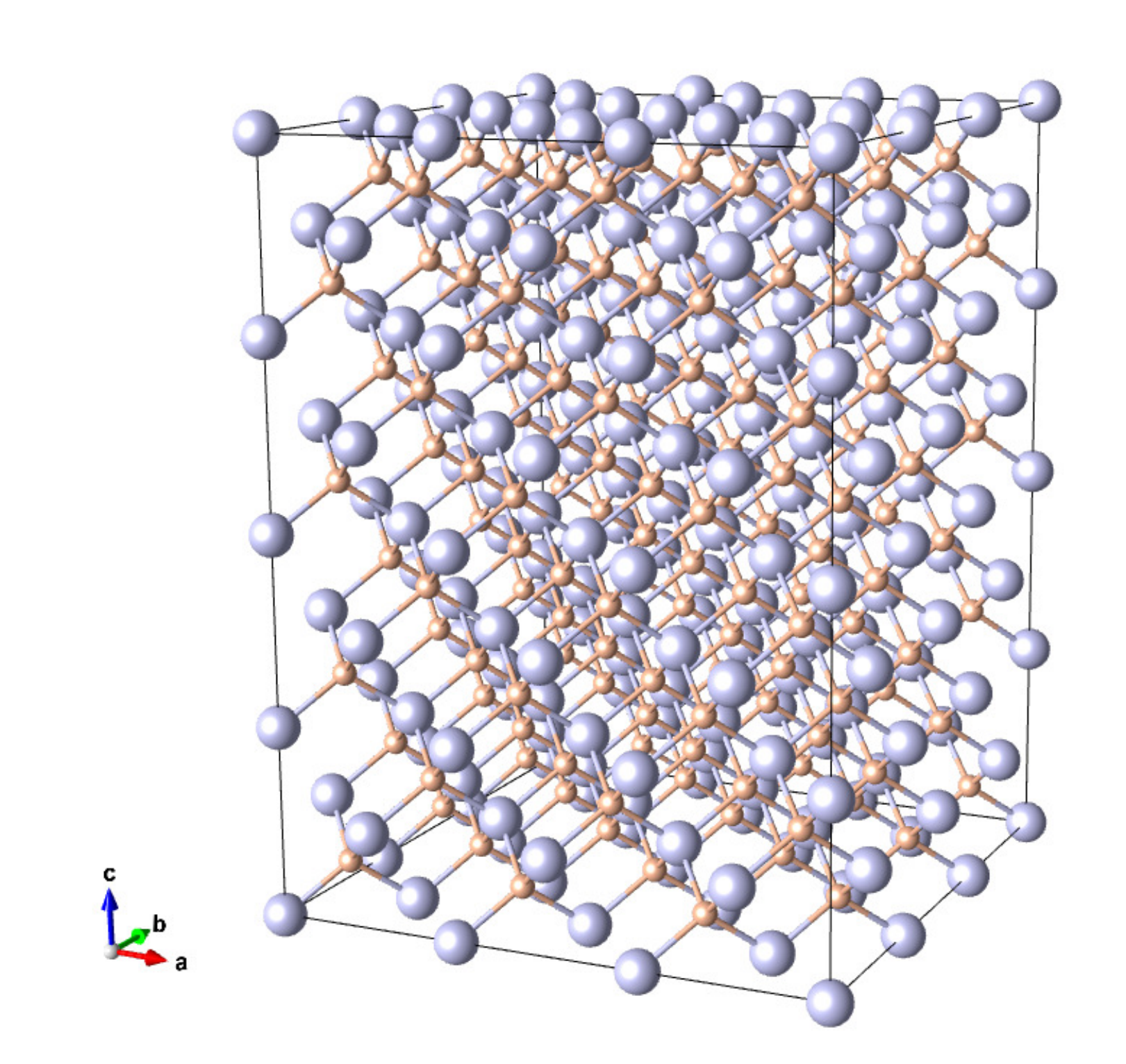}}
\caption{(Colour online) Crystal structure of the $3a\times3a\times4a$ 72-atom zinc-blende supercell. The cation and anion atoms are represented by big indigo and small orange balls, respectively.} \label{f1}
\end{figure}

The calculations have been performed using the plane waves pseudopotential approach within the density functional theory (DFT) \cite{dft}, implemented in the Quantum-ESPRESSO package \cite{qe1,qe2}. The exchange-correlation potential was treated with the generalized gradient approximation (GGA) \cite{gga} of Perdew, Burke, and Ernzerhof (PBE) \cite{pbe}. The projector augmented-wave (PAW) method \cite{paw} was used with the ultrasoft pseudopotentials without linear core correction. The valence shells were distinguished according to the following configuration; [Ar]4$s^{1}$ for potassium (K), [Ar]4$s^{2}$ for calcium (Ca) and [He]2$s^{2}$ 2$p^{3}$ for nitrogen (N). The plane-wave cutoff was taken up to 45 Ry after a convergence check, where the Gaussian smearing technique was used due to a metallic character. The Brillouin zone (BZ) was integrated using the Monkhorste-Pack scheme \cite{ibz} with a 4$\times$4$\times$3 mesh. To model the K$_{x}$Ca$_{1-x}$N alloy, we construct a $3a\times3a\times4a$ 72-atom supercell shown in figure \ref{f1}. We place K and Ca atoms on the ideal cation sites with the compositions $x$ and 1-$x$, respectively, and place N atoms on the ideal anion sites.

\section{Results and discussion}
\subsection{Structural properties}
\begin{figure}[h]
\centerline{\includegraphics[width=0.6\textwidth]{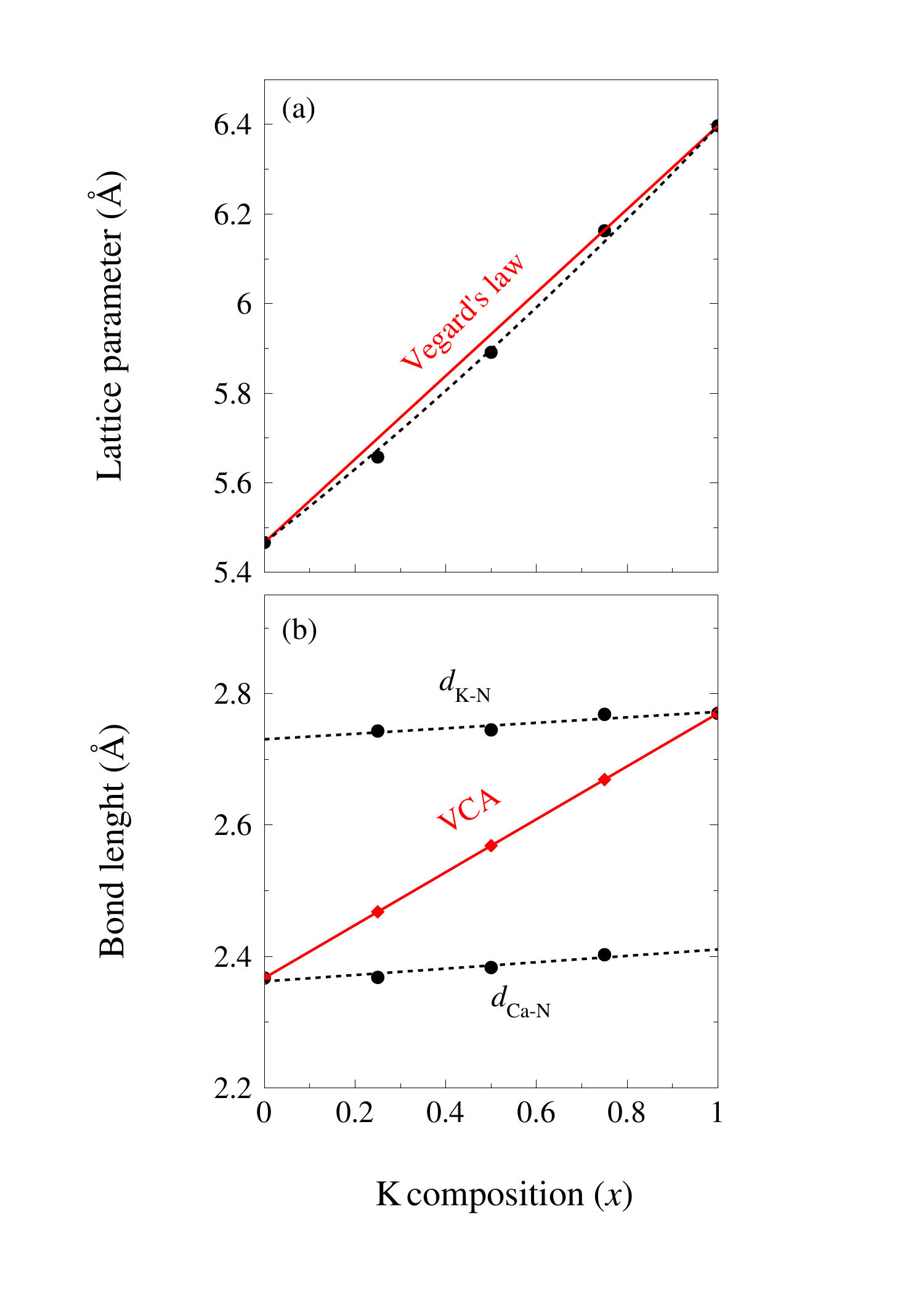}}
\caption{(Colour online) a) The variation of the calculated lattice parameter $a$ as a function of K-composition $x$ in the K$_{x}$Ca$_{1-x}$N alloy (in black closed circle) along with the Vergard's law variation (in red line). b) The average K--N and Ca--N bond lengths as as a function of K-composition $x$ (in black closed circle) along with the VCA estimated bond lengths (in red line).} \label{f2}
\end{figure}

All the results presented in this paper were carried out according to the ferromagnetic order. The magnetic properties were discussed in our previews paper \cite{alloy2}. This section is focused on the structural properties of the ternary alloy K$_{x}$Ca$_{1-x}$N. To study the variation of the lattice parameter and the bond lengths between N and K and Ca atoms of the K$_{x}$Ca$_{1-x}$N alloy, we vary the K composition from $x$ = 0 to 1. The values $x$ = 0 and 1 correspond to the pure CaN and KN, respectively. The lattice parameter was calculated by optimizing the total energy for $x$ = 0, 0.25, 0.5, 0.75, and 1, and fitted with the Vegard's law \cite{Vegard} using the following equation;
\begin{align}
\label{vegard1}
a_{(\textrm{K$_{x}$Ca$_{1-x}$N})} = a_{(\textrm{Vegard})}-\delta_{a}x(1-x),
\end{align}
where
\begin{align}
\label{vegard2}
a_{(\textrm{Vegard})} = xa_{(\textrm{KN})}+(1-x)a_{(\textrm{CaN})}
\end{align}
and $\delta_{a}$ is the deviation parameter. The results are shown in figure \ref{f2} (a). We can see an increase of the lattice parameter $a$ with increasing K-composition $x$ going from CaN to KN. A deviation from the Vegard's law is also observed. Considering the fact that the bonds tend to maintain their natural length, a strong  distortion of the bond takes place. Therefore, a deviation from the Vegard's law is observed and estimated by $\delta_{a}=0.4$~{\AA}. We can attribute this deviation to the structural relaxation effects that are not taken into account by the Vegard's law. These effects are related essentially to the lattice mismatch of about $\sim$16\% between the pure parent compounds. Figure \ref{f2} (b) shows the variation of the K--N and Ca--N bond lengths as a function of the K composition. According to the VCA approximation, the bond lengths must be identical and vary linearly with composition $x$ as shown in figure \ref{f2} (b) (red solid line). As shown in the figure (dashed black lines), both bond lengths increase with increasing K composition. Moreover, the lengths of the bonds are different and close to those of the parent compounds compared to those predicted by the VCA. This can be explained by the fact that each nitrogen atom is surrounded by a mixture of K and Ca atoms, thus defining a shell with two different bonds, K--N and Ca--N.
\subsection{Phase stability}
\begin{figure}[ht]
\centerline{\includegraphics[width=0.6\textwidth]{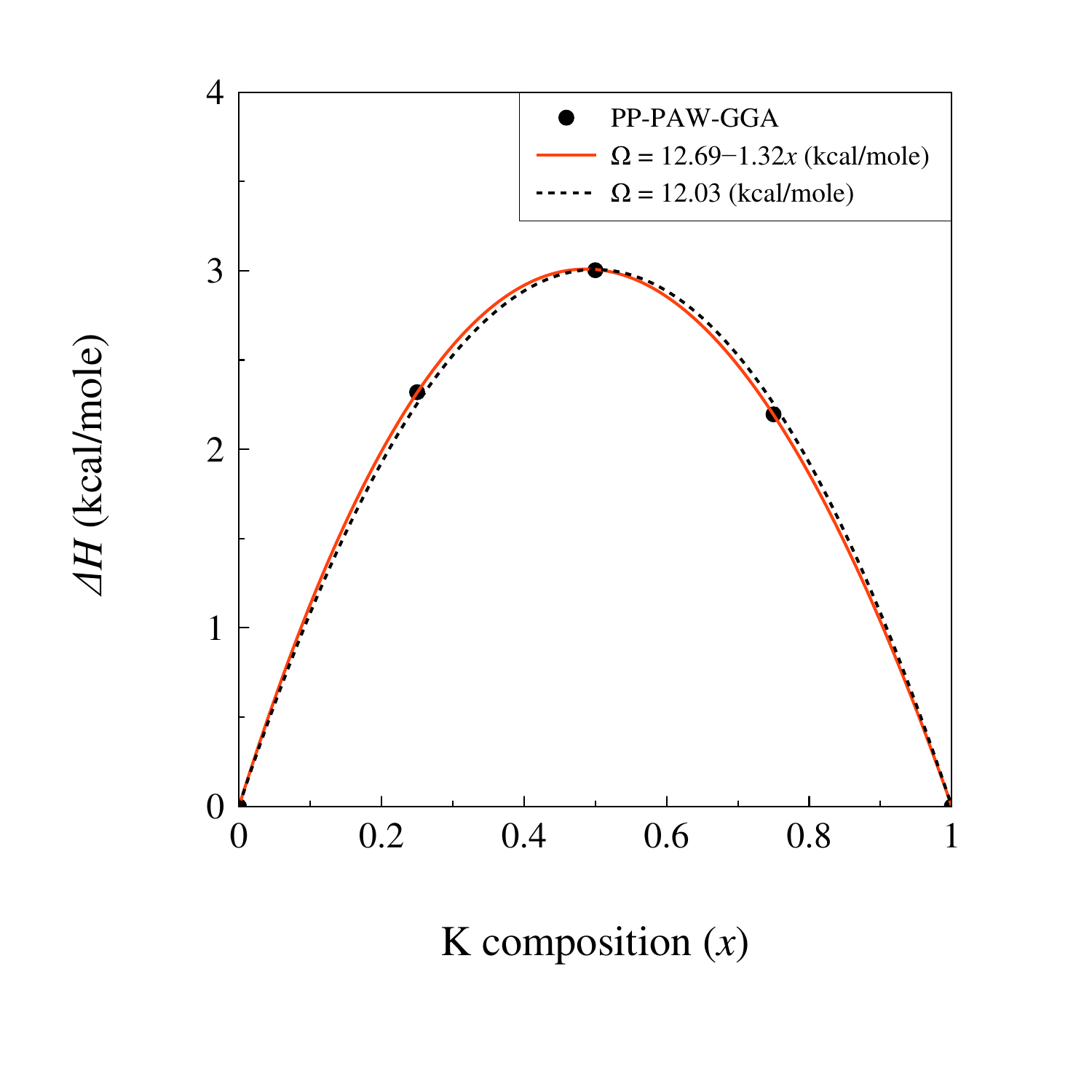}}
\caption{(Colour online) Enthalpy of mixing, $\Delta H$, as a function of the composition $x$ in the K$_{x}$Ca$_{1-x}$N alloy. The results from pseudopotential PBE are shown in closed black circles. The red solid and dashed black curves are calculated using the regular-solution model with the $x$-dependent and $x$-independent interaction parameters $\Omega$, respectively.} \label{f3}
\end{figure}

The study of the phase stability of the K$_{x}$Ca$_{1-x}$N alloy at different K composition is based on the regular-solution model \cite{Regular1,Regular2,Regular3} to predict the pattern of the phase separation of the two partially miscible components. The model assumes that the entropy of mixing is the same as the ideal mixing but the enthalpy of mixing is not zero, so the Gibbs free energy can be expressed by;
\begin{equation}
\label{Gibbs}
\Delta G = \Delta H - T\Delta S,
\end{equation}
where
\begin{equation}
\label{omega1}
\Delta H = \Omega x(1-x)
\end{equation}
is the enthalpy of formation which represents the non-ideality of the mixture. $\Delta S = -R[x\ln(x)+(1-x)\ln(1-x)]$ is the ideal configurational entropy of mixing. $R$ and $T$ are the gas constant and the absolute temperature, respectively. $\Omega$ is the interaction parameter which can depend on $x$. In the regular-solution model, $\Omega$ is the unique material-dependent parameter and can be obtained by expressing the enthalpy of formation $\Delta H$ as the difference in energy between the K$_{x}$Ca$_{1-x}$N and the weighted sum of the constituents:
\begin{equation}
\label{enthalpy}
\Delta H = E_{\text{tot}}(\mathrm{K}_{x}\mathrm{Ca}_{1-x}\mathrm{N}) - xE_{\text{tot}}(\mathrm{KN}) - (1-x)E_{\text{tot}}(\mathrm{CaN}).
\end{equation}
The energy values correspond to the ferromagnetic state and minimized with respect to the lattice parameter. $x$ is the K composition, which takes on the values 0, 0.25, 0.5, 0.75, and 1. To ensure accuracy and comparable $k$-sum convergence, all energy terms in equation (\ref{enthalpy}) are carried out in the same 72-atom supercell. The data are fitted through equation (\ref{omega1}) giving a value of $\Omega=12.03$~kcal/mole. According to the ideal regular-solution model, $\Omega$ is assumed to be $x$-independent \cite{String74} and depends only on the material, but we observe a weak $x$-dependence \cite{Saito99} which cannot be described by equation (\ref{omega1}). Rather, a composition-dependent $\Omega$ should be considered as:
\begin{equation}
\label{omega2}
\Delta H = (\alpha+\beta x)x(1-x),
\end{equation}
where $\Omega$ is given by $\alpha+\beta x$. The best fit gives the $x$-dependent $\Omega=12.69-1.32x$~kcal/mole. The results are shown in figure \ref{f3}. It is clear that the $x$-independent $\Omega$ curve (dashed line) is symmetric about $x=0.5$ while the $x$-dependent $\Omega$ curve (solid line) is a little asymmetric towards a lower $x$ side. This asymmetry on $\Delta H$ curve induces a deviation from the regular solution behaviour that affects the phase diagram. In the range 0$\leqslant x \leqslant$1, $\Omega$ takes on an averaged value, which corresponds to $\Omega=12.69-1.32\times0.5=12.03$~kcal/mole.
\begin{figure}[h]
\centerline{\includegraphics[width=0.6\textwidth]{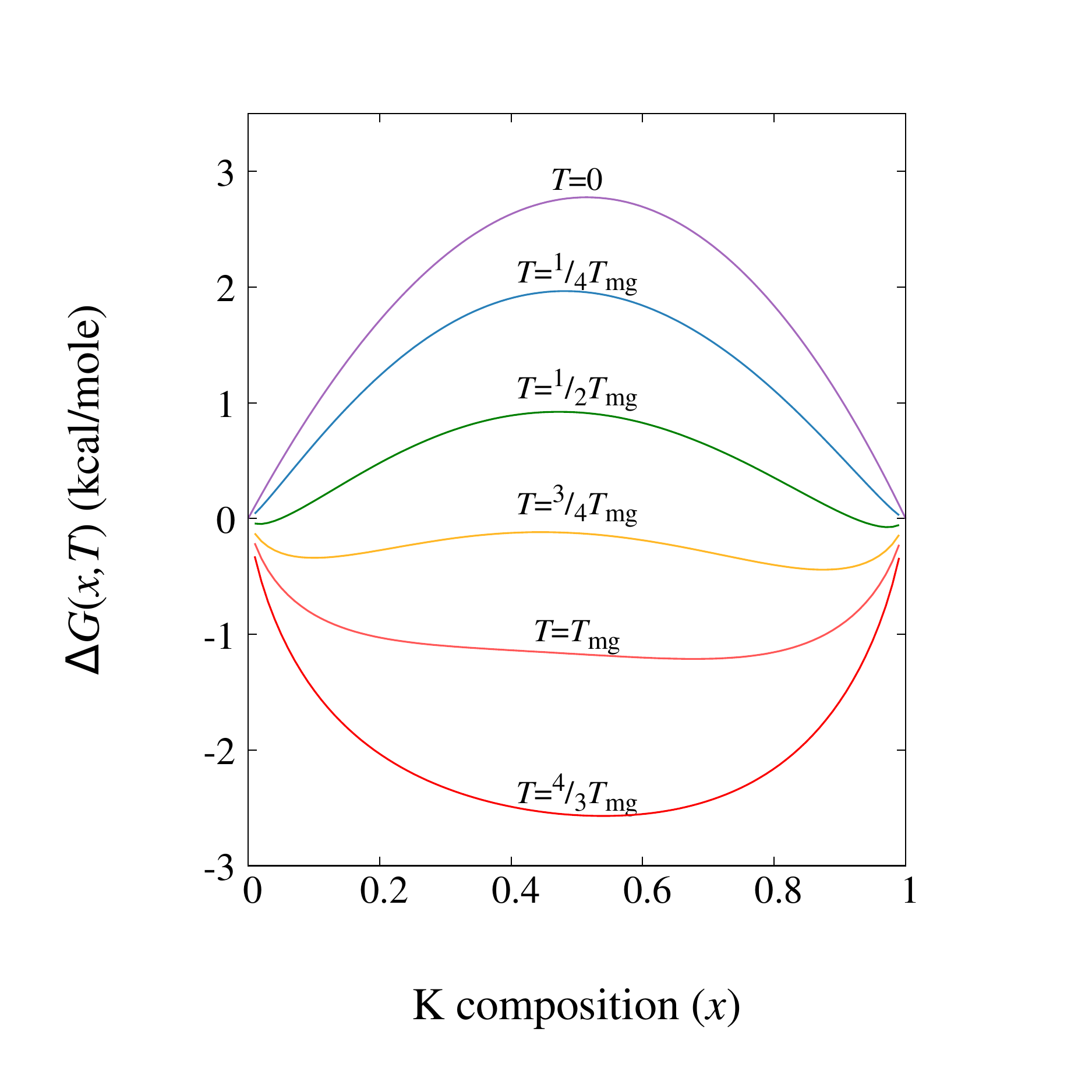}}
\caption{(Colour online) The free energy $\Delta G$ of the K$_{x}$Ca$_{1-x}$N alloy as a function of K composition at some selected values of temperature $T$.} 
\label{f4}
\end{figure}
\begin{figure}[h]
\centerline{\includegraphics[width=0.6\textwidth]{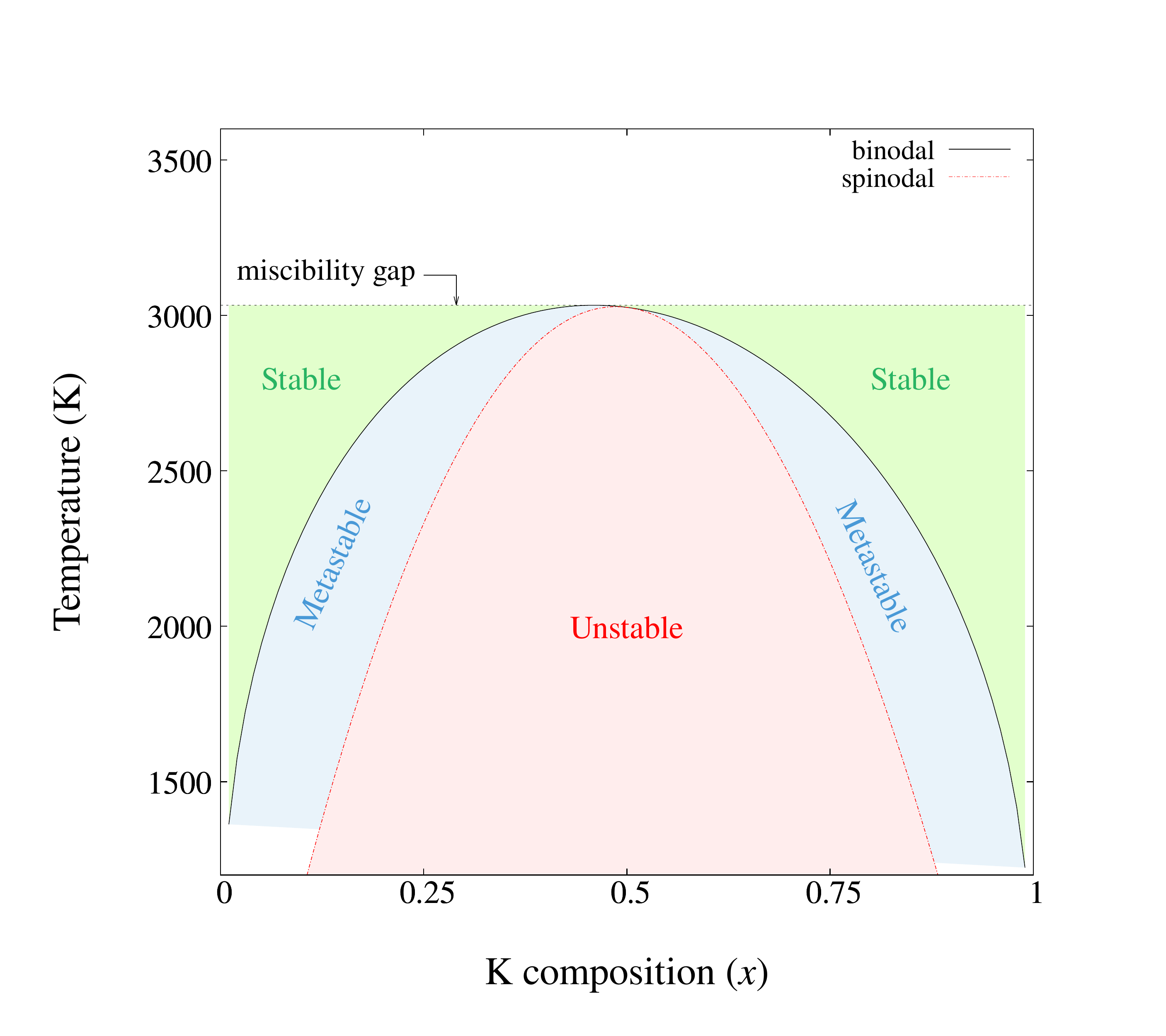}}
\caption{(Colour online) Phase diagram for the K$_{x}$Ca$_{1-x}$N alloy calculated from the regular-solution model using $\Omega=12.69-1.32x$~kcal/mole. The binodal and spinodal curves are shown by the black and red curves, respectively.} 
\label{f5}
\end{figure}

Before investigating the phase stability, one must calculate the Gibbs free energy $\Delta G$ of the K$_{x}$Ca$_{1-x}$N alloy using equation (\ref{Gibbs}) with $\Omega=12.69-1.32x$~kcal/mole obtained above. Figure~\ref{f4} shows the variation of the free energy $\Delta G$ as a function of K composition at some selected values of temperature~$T$. The asymmetry is also observed on different curves as for $\Delta H$. At low temperature, $\Delta G$ exhibits a maximum at $x_c=0.46$ with two minima situated on each side. This suggests the existence of two equilibrium phases with a similar structure at different K composition. By increasing temperature, the two minima converge towards a single minimum at $x_c=0.46$. The critical temperature at which the shape of the $\Delta G$($x$,$T$) curve changes determines the miscibility gap temperature ($T_{mg}$) which corresponds to $\Omega /2k_{\text{B}}\approx$3033 K. From the calculated free energy one can deduce the phase diagram of the K$_{x}$Ca$_{1-x}$N mixing, which separates the one-phase region from the two-phase region in the plane of composition and temperature.

Figure \ref{f5} shows the phase diagram based on the calculated interaction parameter $\Omega=12.69-1.32x$~kcal/mole which shows the stable, metastable and unstable mixing regions of the K$_{x}$Ca$_{1-x}$N alloy. The binodal curve (solid line in figure~\ref{f5}) is found by the common tangent construction in $\Delta G(x,T)$. The spinodal curve (dashed line) determines the equilibrium solubility limit, known as the miscibility gap, and is determined as those points at which the second derivative of $\Delta G$ is zero, i.e., $\partial^{2} G/\partial x^{2}=0$. At temperatures above the binodal curve (green filled region), the alloy is stable and completely miscible within a single phase. Inside the spinodal region (red filled region in figure \ref{f5}), where $\Delta G$ has a negative curvature, the alloy is unstable and two metastable regions (blue filled regions) may exist. At points under the miscibility gap, the phase compositions vary with $T$. We must keep in mind that no short-range ordering or vibrational effects were included in the phase-diagram calculation. The goal here is the effect of the alloy mixing enthalpies obtained from first principles. The existence of a miscibility gap indicates the spinodal decomposition. As shown in figure \ref{f5}, the phase diagram is also asymmetric about $x=0.46$ due to the $x$-dependent interaction parameter $\Omega$.

\section{Conclusion}
In summary, the stable and unstable mixing regions of the K$_{x}$Ca$_{1-x}$N ternary were predicted using the regular-solution model. The calculated interaction parameter depends weakly on K-composition. The curve of the enthalpy as function of the K-composition shows an asymmetry on the left side of $x=0.5$. The critical temperature for K$_{x}$Ca$_{1-x}$N alloy was obtained from the phase diagram and approaches 3033~K, which leads to a wide range of phase separation for typical growth temperatures.


\ukrainianpart

\title{Фазова стійкість та структурні властивості нового феромагнітного сплаву K$_{x}$Ca$_{1-x}$N з перших принципів}
\author{К. Ларбауї,	А. Лакджа,	Г. Бассу}
\address{
	Лабораторія мікроскопії, мікроаналізу речовини та молекулярної спектроскопії,
	університет Джиллалі Ліабес, 22000 Сіді-Бель-Аббес, Алжир
}
%
%
%

\makeukrtitle

\begin{abstract}
	\tolerance=3000%
	Ми вивчаємо структурні властивості та фазову стійкість cплаву K$_{x}$Ca$_{1-x}$N з використанням моделі регулярного розчину на основі повної енергії змішування. Застосовувався псевдопотенціальний підхід разом з функціоналом Пердью, Берка та Ернцергофа (PBE). Ми досліджували розподіл довжин зв'язків в залежності від концентрації~$x$. Було передбачено фазове розшарування двох частково змішуваних компонент та розраховано ентальпію $\Delta H$ з використанням параметра взаємодії $\Omega$. Спостерігається асиметрія на фазовій діаграмі в околі $x=0.46$, зумовлена $x$-залежним параметром взаємодії $\Omega=12.69-1.32x$~ккал/моль. Знайдено рівноважну межу розчинності, відому як щілина змішуваності, яка становить близько 3033~K.
	\keywords лужноземельні метали, розрахунки з перших принципів, фазові діаграми
	
\end{abstract}

\lastpage
\end{document}